\documentstyle[12pt]{article}

\begin{document}

\hfill\vbox{
  \hbox{OHSTPY-HEP-T-96-020}
  \hbox{hep-ph/9608381}
  \hbox{August 1996}
}\bigskip

\begin{center}
{\large\bf HIGH-TEMPERATURE THERMODYNAMICS OF QCD}~\footnote{
Invited talk presented at Workshop on QCD in Paris, June 1996.}
\bigskip

A. NIETO
\bigskip

{\it Department of Physics,
The Ohio State University, Columbus, OH 43210, USA}

\end{center}

\begin{abstract}
Effective-field-theory methods are used to study the
high $T$ limit of QCD. These methods unravel the
contributions to the free energy of QCD at high
temperature from the scales $T$, $gT$, and $g^2T$.
The free energy is explicitly
computed to order $g^5$. Implications for the
application of perturbative QCD to the quark-gluon
plasma are also discussed.
\end{abstract}
\section{Introduction}

Hadronic matter is expected to undergo a phase
transition when $T\sim T_c\simeq 200$~MeV between a low
temperature phase in which it is confined in the form of
hadrons and a high temperature phase in which quarks and
gluons are deconfined.  The latter phase, know as {\em
Quark-Gluon Plasma} (QGP), can be studied by perturbative
methods since the strong coupling constant $\alpha_s$ is
expected to be small at high temperature.

A field theory at high temperature may be described by a
theory in 3 dimensions resulting from decoupling the modes
with zero Matsubara
frequency~\cite{appelquist-pisarski,nadkarni-1,gpy}.  This
dimensionally reduced theory can be interpreted as an
effective field theory whose parameters are to be computed
as a perturbative expansion in the coupling constant of the
original theory~\cite{bn3}.

The effective theory approach is now a well-established
perturbative method to study field theories at finite
temperature~\cite{bn3,bn4,bn5}. It is a useful tool for
organizing calculations of high-order terms in the
perturbation expansion and it is especially powerful in
dealing with non-Abelian gauge theories. It is also an
important conceptual tool that explicitly separates the
different energy scales.  Here I present its application to
the calculation up to order $g^5$ of the free
energy~\cite{bn5} and use the separation of scales to
analyze the convergence of the series~\cite{bn4,bn5}.

\section{The free Energy of QCD to order $g^5$}

The partition function of QCD at finite temperature is
obtained from the Lagrangian $ {\cal L}_{\rm QCD} = (1/ 4)
G_{\mu\nu}^a G_{\mu\nu}^a + \overline{q}\gamma_\mu D_{\mu}
q$, with gauge coupling constant $g$. We consider the
effective theory that results from integrating out the
nonstatic modes. Such a theory, which is called
Electrostatic QCD (EQCD), is made out of only the static
bosonic modes and 3 dimensional.  The free energy density is
$ F_{\rm QCD} = T \left( f_E - \log {\cal Z}_{\rm EQCD}/ V
\right) $, where ${\cal Z}_{\rm EQCD}$ is the partition
function of EQCD obtained from the Lagrangian
\begin{equation}
  {\cal L}_{\rm EQCD} = {1\over 4} G_{ij}^a G_{ij}^a +
    {1\over 2} (D_i A_0)^2 + {1\over 2} m_E^2 A_0^2 +
    {1\over 8} \lambda A_0^4 + \delta{\cal L}_{\rm EQCD}
\end{equation}
whith effective gauge coupling constant $g_E$.  $\delta{\cal
L}_{\rm EQCD}$ represents an infinite series of
non-renormalizable terms. The effects of the fermions are
incorporated into the effective parameters.  By power
counting, the effective parameters at leading order are:
$f_E\sim T^3$, $m_E^2\sim g^2 T^2$, $g_E^2\sim g^2 T$, and
$\lambda_E\sim g^4 T$.

The magnetostatic fields remain massless; therefore, we can
go further in separating the different scales of QCD at high
temperature by integrating out $A_0$. We obtain an effective
theory of EQCD which is called magnetostatic QCD (MQCD). The
free energy density of QCD can be written as $F = T ( f_E + f_M
+f_G)$, where $f_G = - \log {\cal Z}_{\rm MQCD}/ V$, and
${\cal Z}_{\rm MQCD}$ is the partition function of MQCD
obtained from the Lagrangian $ {\cal L}_{\rm MQCD} = (1/4)
G_{ij}^a G_{ij}^a + \delta{\cal L}_{\rm MQCD}\, .  $ The
gauge coupling constant is $g_M$ and $\delta{\cal L}_{\rm
MQCD}$ represents an infinite series of non-renormalizable
terms.

Again, we can use power counting to identify the order of
the leading contribution to the MQCD parameters: $f_M\sim
m_E^3\sim g^3 T^3$ and $g_M^2\sim g_E\sim g^2 T$.
$\lambda_E$ contributes to the free energy only at order
$g^6$; so, if we are interested in the free energy at lower
order, we can ignore $\lambda_E$. Similarly the
non-renormalizable terms of EQCD can also be omitted.

The leading contribution to $f_G$ can be obtained by
realizing that the only parameter with dimensions involved
in MQCD is $g_M$ and therefore the leading contribution is
of order $g_M^6\sim g^6 T^3$. Since we are interested in
computing the free energy of QCD up to order $g^5$ we can
ignore the contribution from $f_G$; it will be enough to
compute $f_E$ and $f_M$. We conclude that the free energy of
QCD up to order $g^5$ is given by $F = T[ f_E(T,g;\Lambda_E)
+ f_M(m^2_E, g_E;\Lambda_E)]$, where $\Lambda_E$ is a
factorization scale that separates the scales $T$ and $gT$.
Therefore, we have to determine $f_E$, $m_E^2$, $g_E$, and
$f_M$. Calculations for $SU(N)$ gauge theory with $n_f$
fermions and results in analytical form are detailed
in~\cite{bn5}; here, I will just state the results in
numerical form for $SU(3)$ with 3 fermions.

The effective mass $m_E$ is computed by matching the
electrostatic screening mass for QCD and EQCD.
\begin{equation}
m_E^2 =  0.93\; {g^2(\mu)\over 4\pi}\; T^2
\left[ 1 +
\left( -\; 0.26 + 1.43\; \log {\mu \over 2 \pi T} \right)
         {g^2 \over 4\pi}
\right] \, .
\label{mE}
\end{equation}
At this order in $g^2$, there is no dependence on the
factorization scale $\Lambda_E$.

The parameter $f_E$ is determined by calculating the
free energy in both full QCD and EQCD, and matching the two
results.
\begin{eqnarray}
  f_E(\Lambda_E) &=& - 14.8\; T^3\; \left[
    1 -\; 0.90\; {g^2 \over 4\pi} \; + \right.
\nonumber \\
  &+& \left. \left(
    18.3 + 6.91\;\log{\Lambda_E \over 2 \pi T} -
    1.30\;\log{\mu \over 2 \pi T} \right)
      \left( {g^2 \over 4\pi} \right)^2
\right] \, ,
\label{fE}
\end{eqnarray}
where $g$ is the coupling constant in the $\overline{\rm
MS}$ renormalization scheme at the scale $\mu$.  We have
used the renormalization group equation of the coupling
constant to shift the scale of the running coupling constant
to an arbitrary renormalization scale $\mu$.

Through order $g^5$, $f_M$ is proportional to the logarithm
of the partition function for EQCD: $f_M \;=\; - \log {\cal
Z}_{\rm EQCD}/ V $.  Now, we have to consider the
cotnribution to $\log {\cal Z}_{\rm EQCD}$ of orders $g^3$,
$g^4$, and $g^5$ which are given by the sum of 1-loop,
2-loop, and 3-loop diagrams.  The details of this
calculation can be found in~\cite{bn5}; the final result is
\begin{equation}
  f_M(\Lambda_E) = - 0.21 m_E^3 \left[
    1 - \left(0.54 + 0.72 \log{\Lambda_E\over 2m_E}\right)
      \left( {g_E^2 \over m_E}\right) -
    0.70 \left( {g_E^2 \over m_E}\right)^2
\right] \, .
\label{fM}
\end{equation}

The coefficient $f_M$ in (\ref{fM}) can be expanded in
powers of $g$ by setting $g_E^2 = g^2T$ and by substituting
the expression (\ref{mE}) for $m_E^2$.  The complete free
energy to order $g^5$ is then $F = (f_E + f_M) T$; in
agreement with the result obtained independently by
Kastening and Zhai~\cite{kastening-zhai}. Note that the
dependence on the arbitrary factorization scale $\Lambda_E$
cancels between $f_E$ and $f_M$, up to corrections that are
higher order in $g$, leaving a logarithm of $T/m_E$.

\section{Convergence of Perturbation Theory}

We have calculated the free energy as a perturbation
expansion in powers of $g$ to order $g^5$.  In this section,
we examine the convergence of the series and it is based
on~\cite{bn4,bn5}. We now ask how small
$\alpha_s\equiv g^2/(4\pi)$ must be in order for the
perturbation expansion to be well-behaved in the sense that
the term of order $n$ is smaller than the term of order
$n-1$. If the series is apparently convergent in this sense,
then it can plausibly be used to evaluate the free energy.
For simplicity, we consider the case $n_f=3$.  If we choose
the renormalization scale $\mu = 2 \pi T$ which is the mass
of the lightest nonstatic mode, the correction to the
leading order result is a multiplicative factor $1 - 0.9
\alpha_s + 3.3 \alpha_s^{3/2} + (7.1 + 3.5 \log \alpha_s)
\alpha_s^2 - 20.8 \alpha_s^{5/2}$.  The $\alpha_s^{5/2}$
term is the largest correction unless $\alpha_s(2 \pi T) <
0.16$ which requires $T$ greater than about 15 GeV; at this
value of $T$, the series is $1 - 0.14 + 0.21 - 0.02 - 0.21 +
\cdots$, and its convergence is questionable. Now suposse
the temperature is a few times above the transition
temperature ($\sim 200$ MeV); {\em e.g.\/}, $T\simeq 350$
MeV then $\alpha_s(2 \pi T)\simeq 0.3$ and the series has
the form $1 - 0.27 + 0.54 + 0.26 - 1.03 + \cdots$. It is
clear that the perturbative series is not convergent at {\em
this\/} temperature. If we consider $T\simeq 1$ TeV, then
$\alpha_s(2 \pi T)\simeq 0.07$ and the series is $1 - 0.063
+ 0.061 - 0.011 - 0.027 + \cdots$. However, one has to
realize that at such high temperatures electroweak processes
become important and that we must also take into account the
effects of heavy quarks.

We can go further in our analysis provided that we
have separated the contributions from the scales $T$ and
$gT$.  The free energy is the sum of two
terms: $f_E$ and $f_M$ which contain the contributions of
order $T$ and $gT$ respectively.
The term $f_E$ is given in~(\ref{fE}), when $n_f = 3$
and $\mu=2\pi T$, the correction to the
leading order result form the series $1 - 0.90\alpha_s +
(6.47 - 6.91 \log{\Lambda_E \over 2 \pi T})\alpha_s^2 +
\cdots$. The next-to-leading-order correction to $f_E$ is
independent of $\mu$ and $\Lambda_E$, and is small compared
to the leading-order term provided that $\alpha_s(2\pi T) <
1.1$ which corresponds to $T>67$ MeV.  The
next-to-next-to-leading-order correction can be made small
by adjusting $\Lambda_E$; it vanishes for $\Lambda_E = 5.1
\,\pi T$.  We conclude that the perturbation series for
$f_E$ is well-behaved if the factorization scale $\Lambda_E$
is chosen to be approximately $5 \pi T$.  Whether this
choice is reasonable can only be determined by calculating
other EQCD parameters to higher order to see if the same
choice leads to well-behaved perturbation series.  It is
interesting to study the convergence of the other parameter
of EQCD that we have computed, $m_E^2$, which is given
by~(\ref{mE}) and when $n_f = 3$ and $\mu=2\pi T$, the
correction to the leading order result form the series $1 -
0.26\alpha_s$.  For the next-to-leading-order correction to
$m_E^2$ to be smaller than the leading-order term, we must
have $\alpha_s(2\pi T) < 3.8$, which corresponds to $T>40$
MeV.  Based on these results, we conclude that the
perturbation series for the parameters of EQCD are
well-behaved provided that $\alpha_s(2 \pi T) < 1$, which
corresponds to $T>70$ MeV.

We now consider the convergence of the perturbation series
(\ref{fM}) for $f_M$.  The size of the next-to-leading-order
correction depends on the choice of the factorization scale
$\Lambda_E$.  It is small if $\Lambda_E$ is chosen to be
approximately $m_E$.  The next-to-next-to-leading-order
correction in (\ref{fM}) is independent of any arbitrary
scales.  If $n_f=3$, it is smaller than the leading order
term only if $\alpha_s < 0.17$, which corresponds to $T>2$
GeV.  Thus the perturbation series for $f_M$ is well-behaved
only for temperatures that are much higher than those
required for the parameters of EQCD to have well-behaved
perturbation series.

This analysis indicates that the slow convergence of the
expansion for $F$ in powers of $\sqrt{\alpha_s}$ can be
attributed to the slow convergence of perturbation theory at
the scale $gT$.

\section{Conclusions}

Our explicit calculations were significantly streamlined by
using effective-field-theory methods to reduce every step of
the calculation to one that involves only a single momentum
scale.  They allow us to study the convergence of the
perturbation expansion for thermal QCD.  At the scale $T$,
perturbation corrections can be small only if $T > 70$ MeV.
At the scale $gT$, perturbation corrections can be small
only if $T > 2$ GeV.  Thus, in order to achieve a given
relative accuracy, the temperature must be much larger for
perturbation theory at the scale $gT$ compared to
perturbation theory at the scale $T$.

There is a range of temperatures in which perturbation
theory at the scale $gT$ has broken down, but perturbation
theory at the scale $T$ is reasonably accurate. In this
case, one can still use perturbation theory at the scale $T$
to calculate the parameters in the EQCD Lagrangian. However,
nonperturbative methods, such as lattice simulations of
EQCD, are required to calculate the effects of the smaller
momentum scales $gT$ and $g^2T$. The effective-field-theory
approach provides a dramatic savings in resources for
numerical computation: the effective field theory is 3
dimensional and quarks are integrated out of the theory,
which reduces it to a purely bosonic problem.

\section*{Acknowledgments}
This work was supported in part by the U.S. Department of
Energy, Division of High Energy Physics, under Grant
DE-FG02-91-ER40690. I would like to thank E.~Braaten for
valuable discussions.

\end{document}